\documentclass[10pt,conference]{IEEEtran}
\IEEEoverridecommandlockouts
\usepackage{cite}
\usepackage{amsmath,amssymb,amsfonts}
\usepackage{algorithmic}
\usepackage{graphicx}
\usepackage{textcomp}
\usepackage{xcolor}
\usepackage{multirow}
\usepackage{balance}
\usepackage{booktabs}
\usepackage{array}
\usepackage{caption}
\usepackage{placeins}
\usepackage{comment}
\def\BibTeX{{\rm B\kern-.05em{\sc i\kern-.025em b}\kern-.08em
    T\kern-.1667em\lower.7ex\hbox{E}\kern-.125emX}}
\usepackage[font={small},labelfont=bf]{caption}
\usepackage{float}
\usepackage{microtype}

\begin{document}

\title{How Well Do SAST Tools Work Across Languages? An Empirical Study Using OCPP Implementations}

\makeatletter
\newcommand{\linebreakand}{%
  \end{@IEEEauthorhalign}
  \hfill\mbox{}\par
  \mbox{}\hfill\begin{@IEEEauthorhalign}
}
\makeatother

\author{
\IEEEauthorblockN{
Zachary Wadhams\IEEEauthorrefmark{1},
Emma Sheppard\IEEEauthorrefmark{1}\IEEEauthorrefmark{2},
Dalton Arford\IEEEauthorrefmark{2},
Clemente Izurieta\IEEEauthorrefmark{1}\IEEEauthorrefmark{2}\IEEEauthorrefmark{3},
Ann Marie Reinhold\IEEEauthorrefmark{1}\IEEEauthorrefmark{2}
}

\IEEEauthorblockA{
\IEEEauthorrefmark{1}Gianforte School of Computing,
Montana State University,
Bozeman, MT, USA
}

\IEEEauthorblockA{
\IEEEauthorrefmark{2}Pacific Northwest National Laboratory,
Richland, WA, USA
}

\IEEEauthorblockA{
\IEEEauthorrefmark{3}Idaho National Laboratory
}
}

\maketitle

\begin{abstract}
As electric vehicle (EV) adoption accelerates globally, securing electric vehicle charging station (EVCS) software becomes increasingly critical. The Open Charge Point Protocol (OCPP), widely adopted for EVCS management, places significant responsibility on developers to securely implement specifications that are subject to multiple interpretations. The variety of OCPP implementations across different programming languages provides a unique opportunity to evaluate security tooling in a real-world, multi-language context where implementation variability may introduce vulnerabilities. This study provides the first empirical evaluation of Static Application Security Testing (SAST) tool performance across multiple programming languages using OCPP implementations as a real-world test corpus. We applied both comprehensive multi-language tools (SonarQube and Semgrep) and language-specific analyzers to twelve OCPP 1.6 implementations---each written in a distinct programming language---to examine how detection patterns vary across language ecosystems. Total issues ranged from just two in Rust to 1,689 in Java. Mature programming languages like Java yielded substantially more findings than newer languages like Rust, likely because established languages have more developed SAST tool ecosystems rather than inherently worse code quality. These results suggest that developers should consider available SAST tooling when selecting implementation languages for security-critical systems.
\end{abstract}


\section{Introduction}
\label{sec:Introduction}
Over the past decade, global electric vehicle (EV) sales have experienced exponential growth, reaching record numbers in 2023 \cite{b1}, and adoption will continue to accelerate \cite{b2} \cite{b3}. A critical element of this expansion is electric vehicle charging stations (EVCS), which bring logistical, technical, and security challenges \cite{b4}. As EVCS software becomes increasingly integrated with grid and vehicle operations, cybersecurity emerges as a crucial concern \cite{b6}. Compromised systems could disrupt vehicle functionality and grid stability, potentially enabling attackers to trigger widespread outages with profound economic and societal impacts \cite{b25}.

Central to EVCS communication is the Open Charge Point Protocol (OCPP), developed by the Open Charge Alliance (OCA) \cite{b7}. OCPP has become the dominant standard, mandated by the UK's Public Charge Point Regulations 2023 \cite{b8}, the U.S. National Electric Vehicle Infrastructure (NEVI) Program \cite{b9}, and promoted by the EU's Alternative Fuels Infrastructure Regulation (AFIR) \cite{b10}. However, the OCA publishes only the protocol specification without reference implementations, placing the burden of secure implementation on developers who must individually interpret abstract security requirements \cite{b16}. While OCPP 1.6 introduced optional security features such as encryption and authentication, the flexibility inherent in high-level specifications can lead to vulnerabilities stemming from limited guidance and varying security expertise \cite{b7}.

One approach to mitigate these risks is Static Application Security Testing (SAST). SAST tools analyze source code without execution, identifying potential vulnerabilities and code quality issues prior to deployment \cite{b11}. Given the need to address security gaps in OCPP implementations, this study conducts a systematic, SAST-driven investigation across open-source OCPP 1.6 implementations. \textit{Our goal is to quantify how programming languages and their associated tooling ecosystems correspond with issue, weakness, and vulnerability detection.} Specifically, we explore:

\textbf{RQ1}: To what extent do SAST tools detect security and quality issues in open-source OCPP 1.6 implementations across different programming languages?

\textbf{RQ2}: How does the maturity and comprehensiveness of static analysis tool support influence the detection of security and quality issues in open-source OCPP 1.6 implementations?

\section{Related Work}
\label{sec:Related Work}

Research has identified significant security vulnerabilities in OCPP's communication framework. Garofalaki et al. \cite{b12} survey OCPP security from a specification-oriented perspective, highlighting attack vectors including unauthorized access, data integrity violations, and denial-of-service threats. Alcaraz et al. \cite{b13} extend this analysis to smart grid environments, examining man-in-the-middle attacks, energy theft, and unauthorized charging operations that could destabilize power networks.

Despite the potential of static analysis for identifying vulnerabilities, significant barriers limit adoption. Shen et al. \cite{b15} find that only 3 percent of open-source embedded projects use SAST tools beyond compiler warnings. Additionally, large variations in reported vulnerabilities occur from varying tool versions \cite{b24} or vendors \cite{b26, b29}, hampering user confidence \cite{b19}. A systematic review of 89 papers confirms that false positives are the most frequently cited barrier, mentioned in over two-thirds of studies \cite{b19}.

Imparato et al. \cite{b14} demonstrate static analysis effectiveness by comparing tools applied to AUTOSAR automotive components, showing improved code quality and safety. Nguyen-Duc et al. \cite{b23} further demonstrate that combining multiple complementary tools provides more comprehensive vulnerability coverage than any single tool alone.

While existing research addresses OCPP security challenges and SAST adoption barriers separately, no comprehensive evaluation exists of how static analysis tools perform across the multi-language ecosystem that characterizes OCPP development. Most prior evaluations focus on single-language settings, leaving cross-language performance largely unexplored—a critical gap for OCPP developers selecting both programming languages and analysis tools.

\section{Methods}
\label{sec:Methods}
Our study required selecting open-source OCPP implementations suitable for examining both security and quality issue detection across diverse programming languages (\textbf{RQ1}) and for evaluating how language choice and the maturity and comprehensiveness of static analysis tools affect detection outcomes (\textbf{RQ2}). To achieve this, we established a systematic approach for identifying and validating suitable repositories.

\subsection{Repository Selection Criteria}

\subsubsection{Protocol Version}
To be included in our study, projects were required to implement OCPP 1.6. Although the Open Charge Alliance now recommends OCPP 2.0.1 and, as of 2025, OCPP 2.1 \cite{b7}, OCPP 1.6 remains the most widely adopted version in production charging systems \cite{b18}. While we considered including newer versions, we found only three open-source projects supporting OCPP 2.0.1 and none implementing 2.1. This limited availability reflects the current adoption landscape, where OCPP 1.6 continues to dominate and newer versions are only gradually being adopted \cite{b20} \cite{b21}. Consequently, we required each selected repository to explicitly document support for OCPP 1.6 through README files or other repository-level indicators such as version-labeled folders or filenames. This ensured that our final corpus consisted exclusively of OCPP 1.6 implementations.

\subsubsection{Implementation Completeness} Repositories must contain actual protocol implementation source code rather than only test frameworks, demonstration code, or documentation. This ensured our analysis focused on production codebases that static analysis tools would encounter in real-world scenarios. To further align with production-oriented code, we used the latest release branch for each repository for analysis. 

\subsubsection{Language Diversity} To address \textbf{RQ2}, we actively sought one implementation per programming language encountered to maximize the language diversity of our analysis. When multiple implementations existed in the same language, we selected a single representative that met all other inclusion criteria and excluded the rest. This allowed us to prioritize breadth across languages rather than depth within any individual language.

\subsection{Search and Validation Process}
We conducted a comprehensive search for open-source OCPP 1.6 implementations across multiple programming languages. For each candidate repository, we performed manual validation by examining both the repository documentation and source code structure to verify that it met our inclusion criteria.

This systematic approach yielded 12 implementations, each written in a different programming language, representing the full set of open-source OCPP 1.6 implementations that met our inclusion criteria. We did not consider factors such as project maturity, maintenance status, or community size, as our focus was on code analysis rather than project viability. The selected implementations span both widely used languages (Java, Python, C, C++) and less common ones (Ruby, PHP, Kotlin, Scala), offering sufficient language diversity for our comparative analysis.

A complete list of the selected implementations is provided in Table \ref{tab:tools_by_language}. To ensure safety and prevent potential exploitation of OCPP implementations deployed in operational charging stations, this study does not identify or link to specific repositories. Instead, implementations are referred to only by the programming language in which they are written. This precaution is taken as we will discuss certain vulnerabilities identified as results, and we aim to avoid providing information that could be used to exploit these weaknesses in real-world systems.

\subsection{Static Analysis Tool Selection}
To address both research questions effectively, we adopted a multi-tool approach combining comprehensive multi-language tools with language-specific analysis capabilities. Our tool selection strategy prioritized practical accessibility and broad language coverage over theoretical performance metrics.

\subsubsection{Comprehensive Multi-Language Tools} We selected SonarQube and Semgrep as our primary analysis tools. These enterprise SAST tools are widely recognized in industry, do not typically require expensive licensing requirements, and provide coverage across all 12 programming languages in our study. This selection enables consistent cross-language comparison while reflecting tools that practitioners commonly encounter in real-world development environments.

\subsubsection{Language-Specific Tools} To complement the comprehensive tools and provide deeper language-specific analysis, we identified one additional SAST tool for each programming language. Our selection criteria prioritized availability and compatibility over feature completeness—several promising tools were excluded due to setup complexity or inability to successfully analyze our selected OCPP implementations without crashing. This pragmatic approach reflects the real-world constraints developers face when adopting SAST tools. Table \ref{tab:tools_by_language} provides the list of language-specific tools selected.

\subsubsection{Tool Configuration} All tools were executed using their default configurations to reflect the ``out-of-the-box" experience that most developers encounter when first adopting SAST tools. This enables our approach to provide insights into tool effectiveness without requiring specialized configuration knowledge, making our findings more applicable to typical development scenarios.

\begin{table}[!tb]
\centering
\renewcommand{\arraystretch}{1.1}
\caption{Language-specific static analysis tools used in addition to Semgrep and SonarQube}
\label{tab:tools_by_language}
\begin{tabular}{@{}ll@{}}
\toprule
\textbf{Language} & \textbf{Tool} \\ 
\midrule
C & Cppcheck \\
C\# & Roslyn \\
C++ & Cppcheck \\
Go & Gosec \\
Java & SpotBugs \\
JavaScript & NodeJsScan \\
Kotlin & Detekt \\
PHP & Psalm \\
Python & Bandit \\
Ruby & Brakeman \\
Rust & Cargo-audit \\
Scala & WartRemover \\
\bottomrule
\end{tabular}
\end{table}

\subsubsection{Analysis Scope and Limitations} Each repository was analyzed using all applicable tools (two comprehensive tools and one language-specific tool). We encountered execution challenges that reflect real-world SAST deployment issues. Semgrep could not execute on the C implementation, and both Semgrep and SonarQube returned no findings for certain language combinations (PHP and Scala for Semgrep, Rust for SonarQube) despite successful execution. These outcomes are reported as part of our results. We make no assumptions about finding accuracy and do not validate for false positives or false negatives, focusing instead on raw detection patterns and comparative analysis across languages and tools.

Following this approach, we conducted our comprehensive analysis across all selected repositories and tools. The results of this multi-tool, multi-language analysis provide insights into both the security landscape of OCPP implementations and the practical effectiveness of SAST tools across diverse programming languages.

\section{Results and Discussion}
\label{sec:ResultsDiscussion}
Our comprehensive analysis of twelve open-source OCPP 1.6 implementations using multiple SAST tools revealed significant variations in security and quality issue detection across languages. The findings demonstrate an interplay between language characteristics, tool maturity, developer implementation choices, and the flexibility of the OCPP specification. Each of these factors may contribute to the observed variation in SAST performance across different implementations.

Table II presents an overview of the findings across all analyzed implementations, categorized by tool type. The results show striking disparities in detection patterns, with total finding counts ranging from just two findings for Rust to 1,689 for Java. These variations reflect not only differences in implementation approaches but also the maturity of both the programming languages and their associated static analysis ecosystems.

\begin{table*}[t!]
\centering
\normalsize
\renewcommand{\arraystretch}{1.15}
\label{tool_findings_by_language}
\caption{Tool findings by programming language across multiple SAST tools. This table presents static analysis results for 12 programming languages using Semgrep, SonarQube, and language-specific security tools. Java shows the highest finding count (1,689) while Rust shows the lowest (2), demonstrating significant variation in detection across languages and analysis tools.}
\begin{tabular}{|l|r|r|l|r|}
\hline
\textbf{Language} & \textbf{Semgrep} & \textbf{SonarQube} & \textbf{Language-Specific Tool} & \textbf{Total Findings} \\
\hline
C\# & 9 & 281 & roslyn \hfill 4 & \textbf{294} \\
\hline
C++ & 17 & 99 & cppcheck \hfill 109 & \textbf{225} \\
\hline
C & 0\textsuperscript{a} & 45 & cppcheck \hfill 56 & \textbf{101} \\
\hline
Go & 39 & 134 & gosec \hfill 19 & \textbf{192} \\
\hline
Java & 4 & 1,678 & spotbugs \hfill 7 & \textbf{1,689} \\
\hline
JavaScript & 3 & 293 & nodejsscan \hfill 15 & \textbf{311} \\
\hline
Kotlin & 1 & 729 & detekt \hfill 70 & \textbf{800} \\
\hline
PHP & 0\textsuperscript{b} & 122 & psalm \hfill 53 & \textbf{175} \\
\hline
Python & 1 & 34 & bandit \hfill 641 & \textbf{676} \\
\hline
Ruby & 85 & 4 & brakeman \hfill 14 & \textbf{103} \\
\hline
Rust & 1 & 0\textsuperscript{b} & cargoaudit \hfill 1 & \textbf{2} \\
\hline
Scala & 0\textsuperscript{b} & 9 & wartremover \hfill 83 & \textbf{92} \\
\hline
\textbf{Total} & \textbf{160} & \textbf{3,428} & \hfill \textbf{1,072} & \textbf{4,660} \\
\hline
\end{tabular}
\vspace{0.5em}
\footnotesize

\textsuperscript{a} Semgrep consistently failed on the C project \\
\textsuperscript{b} Tool executed successfully but found no security issues
\end{table*}

\subsection{Security Through Quality}
Of the 4,660 total findings identified across all implementations, five hundred (11\%) were classified as security-related Common Weakness Enumerations (CWEs), while the remaining 4,160 (89\%) represented code quality and maintainability issues. Rather than viewing these as separate categories, it is essential to recognize that quality and security are fundamentally interconnected in software systems \cite{b27}, particularly in safety-critical domains like OCPP implementations.

Poor coding practices can indirectly contribute to security vulnerabilities by making code harder to review, maintain, and secure over time \cite{b22}. In the context of OCPP implementations, quality issues such as duplicate code, poor exception handling, and naming inconsistencies can obscure security-critical logic. This potentially makes it difficult for developers to identify and address vulnerabilities in a timely manner. The prevalence of issues such as dead code, unused variables, and complexity violations across multiple implementations suggests opportunities for improving overall code quality within the OCPP ecosystem, which, if resolved, will strengthen the security posture of these critical infrastructure components.

This quality-security relationship is particularly important given OCPP's role in managing communication between charging stations and central management systems. Code that is difficult to understand, maintain, or modify increases the likelihood that security patches will be incomplete, that new features will introduce vulnerabilities, or that security-critical logic will be compromised during routine maintenance \cite{b28}.

\subsection{Language-Specific Security and Quality Patterns}
The distribution of security-focused versus quality-focused findings varies dramatically by programming language, revealing important patterns about both language characteristics and their associated tooling ecosystems.

C\# and Ruby implementations stand out as having the highest proportion of security-related findings, with 85\% and 96\% of their issues, respectively, classified as genuine security CWEs. The C\# implementation's 251 security findings were dominated by CWE-117 (Improper log neutralization) with 238 instances. Ruby's 99 security findings encompassed a broader range of web application vulnerabilities, including Cross-Site Scripting (CWE-79), path traversal (CWE-22), and HTTP request smuggling (CWE-444).

In contrast, Java and Kotlin implementations showed remarkably low security finding rates, with less than 1\% of their total findings classified as security issues. Java's 1,689 total findings were dominated by code quality concerns, particularly naming conventions and duplicate literals (CWE-398). Similarly, Kotlin's 800 findings were primarily quality-focused. While these issues may appear less critical than direct security vulnerabilities, their volume could reflect accumulated technical debt that hinders secure development over time \cite{b31}.

The C/C++ implementations present an interesting paradox: while containing classic memory-safety risks including buffer overflow vulnerabilities (CWE-120) and null pointer dereferences (CWE-476), these critical security issues represented only 8\% of the 326 total findings. This distribution underscores the importance of prioritizing even numerically small security findings in memory-unsafe languages, as a buffer overflow can have severe exploitation potential in charging infrastructure.

The Go implementation demonstrated a moderate security profile with 30\% of findings classified as security-related, primarily including privilege escalation (CWE-732), weak pseudo-random number generation (CWE-338), and container hardening gaps relevant to containerized OCPP deployments.

\subsection{Tool Performance and Maturity Considerations}
The substantial difference in detection rates between SonarQube (3,428 findings) and Semgrep (160 findings) may reflect multiple factors, including differences in tool design philosophy, rule set focus, and potentially the longer development history of SonarQube (began work in 2007 as opposed to Semgrep starting in 2017). However, the predominance of quality-focused findings in our results suggests that SonarQube's emphasis on code quality detection may be a more significant factor than tool maturity alone. It is also possible that Semgrep’s lower count reflects differences in default rule coverage or the degree of customization typically applied in practice.

Additionally, language-specific tools collectively identified 1,072 findings, though their individual contributions varied considerably. Tools such as Python's Bandit (641 findings) and C/C++'s Cppcheck (109 and 56 findings, respectively) contributed substantially, while others like Java's SpotBugs (7 findings) and C\#'s Roslyn (4 findings) detected comparatively few issues. This variation reflects differences in detection scope among the language-specific tools. Some, like Bandit, focus heavily on security anti-patterns, while others, like SpotBugs, target specific bug patterns rather than the broader code quality issues that dominate SonarQube's output. Despite this variability, these results suggest that a multi-tool approach, combining comprehensive platforms with language-specific analyzers, can complement detection coverage, particularly when the language-specific tool targets a detection category underrepresented by the broader tools.

The disparity in finding counts between comprehensive and language-specific tools is further explained by differences in detection scope. SonarQube and Semgrep analyze for bugs, vulnerabilities, and code quality issues, whereas many language-specific tools target a narrower category. This contrast is most apparent in Java, where SonarQube identified 1,678 findings that were predominantly code quality and style issues, while SpotBugs, which focuses on bug patterns, found only 7. Conversely, Python demonstrates the opposite pattern. Bandit's extensive security-focused rule set produced 641 findings compared to SonarQube's 34, indicating that Bandit's security detection scope exceeds SonarQube's default Python coverage in that category. These contrasting cases illustrate that disparities in raw finding counts between tools reflect differences in what the tools are designed to detect rather than differences in tool effectiveness.

The influence of language maturity is equally apparent in our results. Established languages like Java (first released in 1995) and C (1970s) show extensive tool support and high detection counts, while newer languages like Rust (first stable release in 2015) exhibit minimal findings, potentially reflecting the limited maturity of its static analysis ecosystem. However, this pattern should be interpreted cautiously, as lower finding counts may also reflect better language design for security (as potentially demonstrated by Rust's memory safety features) rather than inadequate tooling alone.

\subsection{Implications of Language Ecosystem on SAST Outcomes}
The observed variations in security and quality findings across open-source OCPP 1.6 implementations underscore a critical consideration for developers and static analysis toolmakers alike: language choice significantly shapes the landscape of available static analysis options. While programming languages themselves may not inherently predispose implementations to security vulnerabilities, the maturity and breadth of their associated tool ecosystems do directly influence the types and quantities of issues detectable by static analysis.

Developers implementing protocols such as OCPP must thoughtfully navigate the trade-offs inherent in language selection. For instance, widely adopted languages such as Java and C exhibit extensive static analysis tool support, as evidenced by the substantial number of quality and security findings identified by tools such as SonarQube, Semgrep, and language-specific analyzers. Conversely, newer languages like Rust, despite their potential security advantages like built-in memory safety, may currently lack equally mature analysis tool support. This limited tooling may leave critical gaps in automated vulnerability and code-quality detection.

Moreover, our findings suggest an actionable insight for toolmakers: there is value in expanding static analysis coverage for emerging programming languages. Though our analysis included only a single Rust project, the notably lower number of findings raises a question regarding whether Rust's apparent resilience stems from superior language design or simply a gap in available analysis tools. Future research and tool development efforts could benefit from a deeper exploration into this area.

Ultimately, these results emphasize that in the broader context of static application security testing (SAST), careful consideration of language-specific analysis capabilities is essential. Developers must weigh the indirect security implications of language choices by assessing the comprehensiveness and performance of available analysis tools. Simultaneously, toolmakers have an opportunity to enhance software security broadly by proactively addressing the tooling gaps that exist within less-established language ecosystems.

\section{Threats to Validity}
\label{sec:Threats to Validity}

\textbf{Internal Validity.} We used default configurations for all static analysis tools to ensure consistency and reflect ``out-of-the-box" behavior; customized configurations would likely yield different detection patterns \cite{b29}. Additionally, our analysis employed specific tool versions at the time of evaluation, and research demonstrates that different versions of the same tool can produce significantly different results \cite{b24}.

\textbf{External Validity.} Our study analyzed one implementation per programming language, limiting generalizability within each language ecosystem. Detected patterns may reflect developer practices or architectural choices rather than language or tooling characteristics. While our focus on OCPP 1.6 provides domain-specific insights, transferability to other protocols depends on structural similarities, though protocols like OCPI follow similar specification-without-implementation patterns.

\textbf{Construct Validity.} Our study measures tool detection patterns rather than accuracy of findings. We do not attempt to determine whether the detected issues correspond to true vulnerabilities or false positives. Observed variations reflect tool behavior rather than actual security differences between implementations. Furthermore, the language-specific tools selected for each language differ in their detection focus. Some prioritize bug patterns, others security anti-patterns, and others code quality. Which limits the direct comparability of raw finding counts across tools and languages.

\section{Conclusion and Future Work}
\label{sec:Conclusion and Future Work}
This study provides the first empirical evaluation of static application security testing (SAST) tool performance across multiple programming languages in the context of OCPP 1.6 implementations. Addressing \textbf{RQ1}, our analysis of twelve open-source implementations using multiple SAST tools revealed substantial variations in detection patterns, with findings ranging from two issues in Rust to 1,689 in Java. This range demonstrates that SAST tools exhibit highly variable detection capabilities across programming environments, identifying a total of 4,660 issues with marked differences in both volume and type of detected problems.

Addressing \textbf{RQ2}, our findings reveal that programming language choice and associated tool ecosystem maturity significantly influence detection outcomes. The dominance of quality-focused findings (89\% of total detections) over direct security vulnerabilities highlights the interconnected nature of code quality and security, particularly in safety-critical domains like electric vehicle charging infrastructure. Languages with mature tooling ecosystems, such as Java and C\#, showed extensive detection capabilities, while newer languages like Rust exhibited minimal findings, reflecting both tool maturity and potentially superior language design for security.

These results extend beyond the EVCS community to software development projects implementing protocol specifications without reference implementations. The differences in SAST tool performance across programming languages emphasize the importance of tool selection based on specific language ecosystems and the value of multi-tool approaches that combine comprehensive platforms with specialized analyzers. For practitioners across all domains, our study demonstrates that understanding SAST tool performance requires consideration of both the capabilities and limitations of available tooling within their chosen development environment.

Future work should explore complementary approaches such as binary analysis to uncover vulnerabilities that source-code analysis tools might miss, potentially providing additional insight into the security landscape of OCPP or similar protocols. Additionally, future studies could benefit from analyzing multiple implementations per programming language to provide more robust confirmation of whether trends observed in this study, such as the limited findings in Rust or the extensive findings in Java, generalize across larger samples and diverse project contexts. Normalizing detection results by code size (e.g., findings per thousand lines of code) would also enable more meaningful cross-language comparisons by controlling for differences in implementation scale. Furthermore, manual validation of a representative subset of findings across tools and languages would help characterize false positive rates and strengthen confidence in the practical significance of reported detection patterns. In sum, root cause analysis is needed to determine the mechanisms driving our results. 

\section*{Acknowledgments}
\label{sec:Acknowledgments}
This research was conducted with the support from the U.S. Department of Homeland Security (DHS) Science and Technology Directorate (S\&T) and the Department of Energy (DOE) Pacific Northwest National Laboratory (PNNL) under contract IAA\# 70RSAT24KPM000022. Any opinions contained herein are those of the author and do not necessarily reflect those of DHS S\&T or DOE PNNL.

\balance

\end{document}